# Application of Multi factor authentication in Internet of Things domain

Udit Gupta
Information Networking Institute
Carnegie Mellon University, Pittsburgh – Pennsylvania, USA
uditg@andrew.cmu.edu

**ABSTRACT:** Authentication forms the gateway to any secure system. Together with integrity, confidentiality and authorization it helps in preventing any sort of intrusions into the system. Up until a few years back password based authentication was the most common form of authentication to any secure network. But with the advent of more sophisticated technologies this form of authentication although still widely used has become insecure. Furthermore, with the rise of 'Internet of Things' where the number of devices would grow manifold it would be infeasible for user to remember innumerable passwords. Therefore, it's important to address this concern by devising ways in which multiple forms of authentication would be required to gain access to any smart devices and at the same time its usability would be high. In this paper, a methodology is discussed as to what kind of authentication mechanisms could be deployed in internet of things (IOT).

1. INTRODUCTION

As internet continues to evolve with time, it is mandatory for security protocols and procedures to update as well so that users and organizations continue to enjoy its benefits without being concerned about any sort of threats. But as number of nodes increases from millions to probably billions the threat of malware, spam and viruses has only increased. Therefore, it's mandatory to build robust systems which can handle intrusion detections as described in [1 and 36] and as a result neutralize any threat arising out of botnets, malware [2] or spam [3]. Furthermore, research has been conducted in the area of data hiding techniques as described in [34 and 35] to reduce the encryption overhead encountered by the system as well as in the domain of preventing eavesdropping by with the help of spread spectrum modulation as mentioned in [37]. Authentication [6] is one such aspect of internet security which if overlooked allows unauthorized users to access systems and steal data. But with attackers gaining access to sophisticated technologies, the need for multi-factor authentication has come up and how it can be deployed in various systems is described in [4, 5].

Authentication to a system has always resulted in overhead for the systems and delay for the end users. Therefore, a lot of effort over the last decade has been undertaken to address this issue [7, 8, 9 and 10]. Internet of things (IOT) [11, 12 and 13] has introduced another dilemma for the developers wherein due to the daily use of smart devices the delay incurred due to authentication must be minimized. In other words the usability of a system must increase if the user intends to use it on a daily basis. The usability [22] of a system is measured using five factors: speed, efficiency, memorability, learnability and user preference. Speed is concerned with how quickly authentication can be accomplished and efficiency is all about how many times system can be at fault to authenticate wrong entity. On the other hand, learnability implies how easy it is to configure the system in order to authenticate oneself and memorability indicates that once learned, how easy it is to authenticate in the system in future. Finally, the user preference takes into account the user's choice of authenticating the system provided it's safe. While authenticating it's important to consider a trade-off between the above mentioned factors.

## 2. MULTI-FACTOR AUTHENTICATION IN INTERNET OF THINGS

Multi-factor authentication [14 and 15] has gained lot of traction in the last few years. Apart from password authentication there are six other types which are used in most practical applications: retina scans [26], security tokens [31], fingerprint recognition [16 and 17], voice recognition [32], facial recognition [18] and gesture based authentication [19 and 20]. This paper will describe how a combination of these authentication mechanisms can be deployed to secure access to smart 'things' and at the same time ease out the process of gaining access for the end user. We'll look at four applications where these mechanisms if deployed will enhance the security framework: smart offices, smart homes, smart airports and smart cars.

**2.1 Smart homes**: A smart home [21 and 23] refers to a home which has an automated system for monitoring temperature, windows, doors, alarms, alerts, etc. Due to this the house owner can remotely control various systems at home via his/her mobile device. Safety is of paramount importance to people staying at home. There exist several security and authentication mechanisms like passwords, hardware tokens and biometric authentication methods. Studies indicate that password based authentication lags on the memorability factor as far as usability of the system is concerned. Hence it would be advisable to have biometric [24 and 25] authentication where members of house would be given access by configuring the system. But in order to enhance system's security it's important to have multiple hierarchy of authentication. Hence a PIN based authentication coupled with fingerprint recognition would enhance security of smart homes so that even if PIN is lost or stolen, a 2nd layer of authentication in the form of fingerprint would not give access to unauthorized users.

**2.2 Smart offices**: Security at work premises have always been of paramount importance to the organizations. Mostly the only secure aspect regarding access to offices has been an identification (ID) card. But it's important to have multiple hierarchy of authentication since the data in these organizations is mostly confidential. A three factor authentication would be sufficient to secure the premises. First step is to issue RFID [29 and 30] to employees which will give them access to the office campus. Next fingerprint recognition to enter the office building would ensure that nobody else apart from authorized employees would get entry. In order to access certain areas in office buildings like labs or rooms, facial recognition software can be deployed which will grant access to only authorized people. Finally, in order to access the virtual private network (VPNs) remotely a security token along with password can be used.

**2.3 Smart cars:** Automobiles have become an important part of everyone's life. Numbers of auto mobiles have increased exponentially in the last few years and this has also brought an increase in the number of theft cases. Since keys or card can be duplicated or lost it is important to authenticate the driver based on his/her identity. Smart cars can have two definitions: either it'll be an autonomous car capable of driving itself or it'll be a vehicle based on artificial intelligence which will result in automation of vehicle but would still need the driver. In the context of this paper, we'll look at the latter case where person controlling the car will be authenticated. Since car will be driven by multiple people a gesture based authentication would provide the added security.

**2.4 Smart airports:** Security at airport right from check-in to boarding a flight is a tedious task due to the fact that it is done manually. Some aspect of airport security can be automated via any of the above mentioned authentication procedure. The ideal way to authenticate a person would be to have a location based authentication wherein the person will be detected at the entrance (via fingerprint or facial recognition) and then will be given access to that area based on the tickets and his identity. Then once the passenger goes for security check-in, fingerprint recognition software would enable the airport authorities to verify his/her identity.

3. **COMPARISONS BETWEEN DIFFERENT AUTHENTICATION MECHANISMS**

Table 1 provides a holistic view about various authentication mechanisms and their usability in terms of four factors: Speed, efficiency, learnability, memorability. We would categorize each of the authentications talked above in terms of these four factors and arrive at the conclusion as to which one is the most efficient. As it can be seen from user's point of view for facial recognition, speed and efficiency are categorized as 'medium' while learnability and memorability is categorized as 'easy'. This is because user doesn't have to remember any of the text in order to authenticate himself while secure system generally takes time for facial recognition and sometimes falters as well in case of any change to facial features. On the other hand in case of Password or PIN based authentication, system would quickly recognize them (provided the search within the database is quick) but user will have to first learn to create and then remember passwords or PIN for a certain time period. Given that numbers of restrictions are imposed on users these days about not keeping common words as their passwords in order to prevent dictionary or brute force attacks [27, 28 and 33], memorability factor further aggravates due to inability of many users to remember them over a time period. Hence, 'speed' and 'efficiency' are termed 'fast' and 'good' respectively but 'learnability' and 'memorability' are 'medium'.

Now let us consider the case for fingerprint recognition. This form of authentication is found to be leading in all the usability factors since user doesn't have to remember anything while system can easily recognize user quickly and efficiently. In 'gesture' based authentication, speed and efficiency of a system to authenticate user would be on higher end but user will have to learn to first configure the system and then remember gesture over a period of time. As a result, the last two factors of usability are termed as 'medium'. Finally, location based authentication would be a combination of multiple forms of authentication discussed previously (mainly fingerprint and facial) since in one of its application (smart airport) user first needs to be identified to be given access to enter and then further checking of that user needs to be done in order to verify his/her identity. Hence this form of authentication may not expedite the process of user authorization but it certainly is the most secure form of authentication in high security zones.

## 4. CONCLUSION

As internet of things continues to gain traction in the field of computing, it has given rise to number of security vulnerabilities. Authentication is one such aspect of the many security issues which was highlighted in this paper and an attempt was made to enable the reader to understand various authentication mechanisms which can be deployed in smart entities (or things) to enhance security. Furthermore, a comparison was made between various authentication (biometric and text based) procedures in the form of table (Table 1) to provide a holistic view to the reader about their usage in IOT.

Table 1

|  | Speed | Efficiency | Learnability | Memorability |
|---|---|---|---|---|
| Facial recognition | Medium | Medium | Easy | Easy |
| Password or PIN | Fast | Good | Medium | Medium |
| Fingerprint | Fast | Good | Easy | Easy |
| Location based | Medium | Medium | Easy | Easy |
| Voice based | Medium | Medium | Easy | Easy |

**REFERENCES**

1. Manoj Rameshchandra Thakur, Sugata Sanyal, "A Multi-Dimensional approach towards Intrusion Detection System", arXiv preprint arXiv:1205.2340, 2012/5/10
2. Manoj Rameshchandra Thakur, Divye Raj Khilnani, Kushagra Gupta, Sandeep Jain, Vineet Agarwal, Suneeta Sane, Sugata Sanyal, Prabhakar S. Dhekne; "Detection and Prevention of Botnets and malware in an enterprise network"; http://arxiv.org/ftp/arxiv/papers/1312/1312.1629.pdf, Volume 5, Issue 2, 2012
3. Zoltan Gyongi, Hector Garcia-Molina, Jan Pedersen, "Combating web spam with trustbank", VLDB '04 Proceedings of the Thirtieth international conference on Very large data bases - Volume 30, Pages 576-587, 2004
4. Ayu Tiwari, Sudip Sanyal, Ajith Abraham, Svein Johan Knapskog, Sugata Sanyal, "A Multifactor Security Protocol For Wireless Payment-Secure Web Authentication using Mobile Devices", arXiv preprint arXiv:1111.3010, 2011/11/13
5. Sugata Sanyal, Ayu Tiwari and Sudip Sanyal, A Multifactor Secure Authentication System for Wireless Payment", Emergent Web Intelligence: Advanced Information Retrieval, Springer London, pages 341-369, 2010/1/1.
6. M. Burrows, M. Abadi and R. M. Needham, "A logic of authentication", University of Cambridge Computer Laboratory, http://www.hpl.hp.com/techreports/Compaq-DEC/SRC-RR-39.pdf

7. Leslie Lamport, "Password authentication with insecure communication", Communications of the ACM, Volume 24 Issue 11, Pages 770-772, Nov 1981
8. Min-Shiang Hwang, Li-Hua Li, "A new remote user authentication scheme using smart cards", IEEE Transactions on Consumer Electronics, Volume 46, Issue 1, Pages 28-30, Feb 2000
9. BC Neuman, T Ts' O, "Kerberos: An authentication service for computer networks", Communications Magazine, IEEE, 1994, http://gost.isi.edu/publications/kerberos-neuman-tso.html
10. S. P. Miller , B. C. Neuman , J. I. Schiller , J. H. Saltzer, "Kerberos authentication and authorization system (1987)", In Project Athena Technical Plan, 1987, http://web.mit.edu/saltzer/www/publications/athenaplan/e.2.1.pdf
11. Feng Xia, Laurence T.Yang, Lizhe Wang and Alexey Vinel, "Internet of Things", INTERNATIONAL JOURNAL OF COMMUNICATION SYSTEMS, Volume 25, Issue 9, pages 1101-1102, September 2012
12. Kevin Ashton, "That 'Internet of Things' Thing", RFID journal, June 2009, http://www.rfidjournal.com/articles/view?4986
13. Kortuem G, Kawsar F Fitton D, Sundramoorthy V, "Smart objects as building blocks for internet of things", Internet Computing IEEE, Volume 14, Issue 1, Pages 44-51, Feb 2010
14. ST Dispensa, "Multi factor authentication", US Patent, Publication number US8365258 B2, Jan 29 2013, http://www.google.co.in/patents/US8365258
15. Abhilasha Bhargav-Spantzel, Anna Squicciarini, Elisa Bertino, "Privacy preserving multi-factor authentication with biometrics", Proceedings of the second ACM workshop on Digital identity management, Pages 63-72, 2006, http://docs.lib.purdue.edu/ccpubs/313/
16. A Jain, L Hong, R Bolle, "On-line fingerprint verification", Pattern Analysis and Machine Intelligence, IEEE Transactions, Volume:19, Issue: 4, Pages 302-314, April 1997
17. AK Jain, S Prabhakar, L Hong, S Pankanti, "Filterbank-based fingerprint matching", Image Processing, IEEE Transactions, Volume:9, Issue: 5, Pages 846-859, May 2000
18. Ashok Samal, Prasana A. Iyengar, "Automatic recognition and analysis of human faces and facial expressions: a survey", Pattern recognition, Volume 25, Issue 1, Jan 1992
19. Shwetak N. Patel, Jeffrey S. Pierce, Gregory D. Abowd, "A gesture-based authentication scheme for untrusted public terminals", Proceedings of the 17th annual ACM symposium on User interface software and technology, Pages 157-160, October 2004
20. Y Niu, H Chen,"Gesture authentication with touch input for mobile devices", Security and Privacy in Mobile Information and Communication Systems, Third International ICST Conference, MobiSec 2011, Aalborg, Denmark, May 17-19, 2011, Revised Selected Papers, pp 13-24
21. Elizabeth Stobert, Robert Biddle, "Authentication in the Home", Workshop on Home Usable Privacy and Security (HUPS) July 24, 2013, Newcastle, UK, http://cups.cs.cmu.edu/soups/2013/HUPS/HUPS13-ElizabethStobert.pdf
22. R. Dhamija and L. Dusseault, "The Seven Flaws of Identity Management: Usability and Security Challenges", Security & Privacy, IEEE, Volume 6, Issue 2, Pages 24-29, April 2008
23. D. Bregman, "Smart Home Intelligence – The eHome that Learns", International Journal of Smart Home, Volume 4, No. 4, October 2010
24. Patrick S. P. Wang, Svetlana N. Yanushkevich, "BIOMETRIC TECHNOLOGIES AND APPLICATIONS", Proceeding (549) Artificial Intelligence and Applications - 2007, http://citeseerx.ist.psu.edu/viewdoc/download?doi=10.1.1.72.3174&rep=rep1&type=pdf

25. G Lawton, "Biometrics: A New Era in Security", IEEE Computer Society, pp.16 - 18, Volume 31, August 1998
26. S.V. Sheela and P.A Vijaya, "Iris Recognition Methods - Survey", International Journal of Computer Applications, Vol. 3, No.5. pp. 19 - 25, June 2010
27. S Delaune, F Jacquemard, "A theory of dictionary attacks and its complexity", Computer Security Foundations Workshop, 2004. 17th IEEE, Pages 2-15, June 2004
28. Lars R. Knudsen, Matthew J. B. Robshaw, "Brute Force Attacks", The Block Cipher Companion Information Security and Cryptography 2011, pp 95-108, 2011
29. A Juels, "RFID security and privacy: A research survey", Selected Areas in Communications, IEEE Journal on Volume: 24, Issue: 2, Pages 381 – 394, Feb 2006
30. R Want, "An introduction to RFID technology", Pervasive Computing, IEEE Volume 5, Issue 1, Pages 25-33, Jan-March 2006
31. F Aloul, S Zahidi, W El-Hajj, "Two factor authentication using mobile phones", Computer Systems and Applications, 2009. AICCSA 2009. IEEE/ACS International Conference, Pages 641-644, 10-13 May 2009
32. L Muda, M Begam, I Elamvazuthi, "Voice recognition algorithms using mel frequency cepstral coefficient (MFCC) and dynamic time warping (DTW) techniques", Journal of Computing, Volume 2, Issue 3, March 2010
33. Vipul Goyal, Virendra Kumar, Mayank Singh, Ajith Abraham and Sugata Sanyal: "A New Protocol to Counter Online Dictionary Attacks", Computers and Security, Volume 25, Issue 2, pages 114-120, Elsevier Advanced Technology, March, 2006.
34. Sandipan Dey, Ajith Abraham, Bijoy Bandyopadhyay and Sugata Sanyal, "Data Hiding Techniques Using Prime and Natural Numbers" Journal of Digital Information Management, arXiv preprint arXiv:1003.3672, ISSN 0972-7272, Volume 6, No 3, pp. 463-485, 2008
35. Sandipan Dey, Hameed Al-Qaheri and Sugata Sanyal, "Embedding Secret Data in HTML Web Page", arXiv preprint arXiv:1004.0459, pp. 474-481. ISBN: 978-83-60434-62-8, 2010/4/3
36. Animesh K Trivedi, Rajan Arora, Rishi Kapoor, Sudip Sanyal and Sugata Sanyal, "A Semi-distributed Reputation-based Intrusion Detection System for Mobile Ad hoc Networks", Journal of Information Assurance and Security (JIAS), Volume 1, Issue 4, December, 2006, pp. 265-274, arXiv preprint arXiv:1006.1956, 2010/6/10
37. Sugata Sanyal, Rohit Bhadauria and Chittabrata Ghosh, "Secure Communication in Cognitive Radio Networks", Computers and Devices for Communication, 2009. CODEC 2009, IEEE, Pages 1-4, 2009/12/14